\documentclass[final,5p,times,twocolumn]{elsarticle}
\usepackage{graphicx}
\usepackage{amssymb}
\usepackage{amsmath}
\usepackage{lineno,hyperref}
\usepackage{color}
\hypersetup{colorlinks=true,linkcolor=red,citecolor=cyan,}

\journal{Physics of the Dark Universe}

\begin{document}

\begin{frontmatter}

%% \title{Title\tnoteref{label1}}
%% \tnotetext[label1]{}
%% \author{Name\corref{cor1}\fnref{label2}}
%% \ead{email address}
%% \ead[url]{home page}
%% \fntext[label2]{}
%% \cortext[cor1]{}
%% \affiliation{organization={},
%%            addressline={}, 
%%            city={},
%%            postcode={}, 
%%            state={},
%%            country={}}
%% \fntext[label3]{}

\title{Gravitational analogue of radiation reaction in STVG}

\author[1,2,3]{Bobur Turimov\corref{cor}}\ead{bturimov@astrin.uz}
\author[3,4]{Abdurakhmon Nosirov}\ead{abdurahmonnosirov000203@gmail.com} 
\author[3,4,5]{Ahmadjon Abdujabbarov}\ead{ahmadjon@astrin.uz}
\address[1]{New Uzbekistan University, Movarounnahr St. 1, Tashkent, 100000, Uzbekistan}
\address[2]{Central Asian University, Milliy Bog St. 264, Tashkent 111221, Uzbekistan}
\address[3]{Ulugh Beg Astronomical Institute, Astronomy St. 33, Tashkent 100052, Uzbekistan}
\address[4]{National University of Uzbekistan, Tashkent 100174, Uzbekistan}
\address[5]{Institute of Fundamental and Applied Research, National Research University TIIAME, Kori Niyoziy 39, Tashkent 100000, Uzbekistan}

\begin{abstract}
We have studied influence of STVG in circular motion of massive particle around the Schwarzschild-MOG black hole and discussed the Innermost Stable Circular Orbit (ISCO) and Marginally Bound Orbit (MBO) position of massive test particle. This research explores dynamics of massive particles within the framework of Scalar-Tensor-Vector Gravity (STVG) when navigating the vicinity of Schwarzschild-MOG black holes. The study extends its investigation to the angular velocity of massive particles orbiting Schwarzschild-MOG black holes, comparing it to predictions from general relativity. Finally, we have investigated gravitational analogue of radiation reaction by considering Landau-Abraham-Dirac equation in STVG. We have shown that because of radiation reaction massive test particle falls down to the black hole.
\end{abstract}

\begin{keyword}
Schwarzschild-MOG spacetime \sep Radiation reaction \sep particle trajectories
\end{keyword}
\end{frontmatter}

\section{Introduction}

Radiation reaction refers to the phenomenon in physics where a charged particle experiences a deceleration or acceleration due to the emission of electromagnetic radiation (photons) as it moves through space. This phenomenon is a consequence of the particle's interaction with its own electromagnetic fields according to classical electrodynamics and when a charged particle accelerates or decelerates, it produces changing electric and magnetic fields, which in turn generate electromagnetic radiation. This radiation carries away energy and momentum from the particle, causing it to lose kinetic energy and slow down. The particle's motion is affected by its own emitted radiation, leading to a self-interaction that can influence its trajectory \cite{Jackson1999AJP,Dirac1938RSPSA,Landau-Lifshitz2}.

The concept of radiation reaction is important in various areas of physics, including radiation theory, high-energy particle physics and astrophysics. For instance, the effect of the radiation reaction force in the trajectory of charged particle around the magnetized Schwarzschild spacetime has been investigated in \cite{Tursunov2018ApJ,Tursunov2018AN,Shoom2015PRD}. On the other hand, it is also interesting to consider the gravitational  radiation reaction in a curved spacetime and in Refs. \cite{Damour1981PLA,Damour1983PRL,Blanchet1986RSPSA,Cutler1994PRD}, radiative gravitational field and gravitational radiation reaction has been extensively studied in the framework of general relativity. 

In our preceding research \cite{TURIMOV2023PLB}, we have suggested new approach that explain gravitational analogue of synchrotron radiation from massive particle orbiting a black hole in Scalar-Tensor-Vector Gravity (STVG) theory which is a modification of General Relativity (GR) that incorporates an additional scalar and vector fields \cite{Moffat2006JCAP,Moffat2015EPJCa}. According to this theory the geodesic equation is not valid for massive particle and its four-acceleration is not equal to zero. It means massive particles generate gravitational radiation. 

In later works MOG field equations were directly obtained from the least action principle and solutions of those equations were compared to cosmological and astrophysical observations \cite{Moffat2006CQG}.
The theory was successfully applied for explanation of galaxy rotation curves \cite{Moffat2008ApJ}, cluster dynamics \cite{Brownstein2007MNRAS}, gravitational lensing \cite{Moffat2007IJMPD,Moffat2009MNRASa,Rahvar2019MNRAS, Moffat2018Galaxies, Ovgun2019AP, Tuleganova2020GRG, Izmailov2019MNRAS}, black hole shadow \cite{Moffat2015EPJC,Moffat2020PRD}, quasi-periodic oscillations \cite{Kolos2020EPJC,Pradhan2019EPJC,Rayimbaev2021Gal} and quasi-normal modes \cite{Manfredi2017JPCS, Manfredi2018PLB,Manfredi2019JURP}. Some other studies on galaxy clusters and acceleration data include \cite{Moffat2019MNRAS, Moffat2019MNRAS, Green2019PDU}. There is a wide range of works on particle motion around compact gravitational objects in MOG focusing on different aspects, namely stable orbits \cite{Sharif2017EPJC,Lee2017EPJC, Perez2017PRD} and energy extraction \cite{Pradhan2019EPJC}. The motion of S2 star orbiting around Sagittarius A* SMBH has been studied in \cite{Monica2022Universe,Monica2022MNRAS,Turimov2022MNRAS,Monica2023MNRAS}. 

The need for modified theories of gravity in astrophysics stems from the perplexing astrophysical phenomena that elude explanation within the confines of traditional general relativity established by Einstein. General relativity has been remarkably successful in describing gravitational interactions at a macroscopic scale~\cite{Moffat2007IJMPD}. However, at the vast reaches of the Universe, significant discrepancies arise, necessitating a reevaluation of our fundamental understanding of gravity. Dark matter and dark energy, proposed to account for unexplained gravitational behaviors, remain elusive. Modified theories, such as $f(R)$ gravity~\cite{Capozziello2020IJMPD,Bajardi2022EPJP} or Modified Gravity~\cite{Moffat2006JCAP,Moffat2009CQG}, offer promising alternatives to bridge these gaps. By modifying the fundamental gravitational principles, these theories seek to elucidate galactic rotation curves, gravitational lensing, and the dynamics of large-scale structures in the universe. A deeper comprehension of gravity on cosmic scales is vital for accurately interpreting astrophysical phenomena and advancing our understanding of the fundamental forces shaping the Universe.

In this paper, we are aimed to investigate of the effect of radiation reaction in dynamical motion of test particle around the Schwarzschild-MOG black hole in STVG. We study gravitational analogue of radiation reaction by considering Landau-Abraham-Dirac equation in STVG. The paper is organized as follows. In Sect.~\ref{Sec:1}, {we discuss circular motion of a massive particle around the Schwarzschild-MOG black hole and present how the innermost stable circular orbit (ISCO) and marginally bound orbit (MBO) for massive particle depend on the MOG parameter.} In Sect.~\ref{Sec:2}, we investigate the gravitational analogue of radiation reaction for massive particle orbits around the Schwarzschild-MOG black hole. Finally, in section~\ref{Sec:3}, we summarize the main obtained results. {Throughout the paper, we adopt a space-like signature $(-,+,+,+)$, a system of units in which $G_N=c=1$.}

\section{Basic equations}\label{Sec:1}

Here we briefly discuss dynamical motion of massive particle around the Schwarzschild-MOG black hole in STVG. In the Boyer–Lindquist coordinates $x^\mu=(t,r,\theta,\phi)$, {the spacetime metric is expressed as~\cite{Moffat2015EPJC} 
\begin{align}
ds^2=-fdt^2+f^{-1}dr^2+r^2(d\theta^2+\sin^2\theta d\phi^2)\ ,
\end{align}
with $f=1-2(1+\alpha)M/r+\alpha(1+\alpha)M^2/r^2$, where $M$ is the total mass of the black hole, $\alpha$ is STVG (MOG) parameter.} The spacetime metric is associated with the following vector potential $\Phi_\mu=-\delta_\mu^t\sqrt{\alpha}M/r$ regarded the fifth force with massive particle. The horizon of the Schwarzschild-MOG black hole is located as $r_H^+=M\left(1+\alpha+\sqrt{1+\alpha}\right)$, which is always greater than the Schwarzschild radius corresponding to the positive values of $\alpha$ parameter. 

According to STVG theory massive particle does not follow geodesic line. Therefore, equation of motion or the four-acceleration of test particle of mass $m$ in Schwarzchild-MOG spacetime is defined as \cite{Moffat2015EPJCa}
\begin{align}\label{eom}
w^\mu=\frac{Du^\mu}{ds}=\frac{\tilde q}{m}B^\mu_{~\nu}u^\nu\ ,\qquad \frac{Du^\mu}{ds}=\frac{du^\mu}{ds}+\Gamma^\mu_{\nu\lambda}u^\nu u^\lambda\ ,
\end{align}
where $u^\mu=dx^\mu/ds$ is the four-velocity of particle normalized as $u_\mu u^\mu=-1$ and orthogonal to the four-acceleration of particle i.e. $u_\mu w^\mu=0$, while $B_{\mu\nu}=\partial_\mu\Phi_\nu-\partial_\nu\Phi_\mu$ is anti-symmetric tensor, and coupling constant ${\tilde q}$ is defined ${\tilde q}=\kappa m=\sqrt{\alpha}m$ {which means MOG parameter always should be positive, i.e. $\alpha\geq 0$}. 

As one can see from equation \eqref{eom} that massive particles do not follow the geodesic line in this theory. Due to the symmetry of the spacetime, there are two constants of motion, namely, specific energy and specific angular momentum i.e. ${\cal E}=f{\dot t}$ and ${\cal L}=r^2\sin^2\theta\,{\dot\phi}$. For simplicity, one can consider circular motion of massive particle in the equatorial plane (i.e. $\theta=\pi/2$), and using normalization of the four-velocity, equation for radial motion in the  Schwarzchild-MOG spacetime yields
\begin{align}\nonumber
\dot r^2&=\left({\cal E}-\frac{\alpha M}{r}\right)^2-f\left(1+\frac{{\cal L}^2}{r^2}\right)\\&=\left[{\cal E}-V_+(r)\right]\left[{\cal E}-V_-(r)\right]\ ,\label{eqr}    
\end{align}
where the effective potential $V_\pm(r)$ is defined as
\begin{align}\label{Veff}
V_\pm(r)=\frac{\alpha M}{r}\pm\sqrt{f\left(1+\frac{{\cal L}^2}{r^2}\right)}\ .
\end{align}
\begin{figure}
\includegraphics[width=8cm]{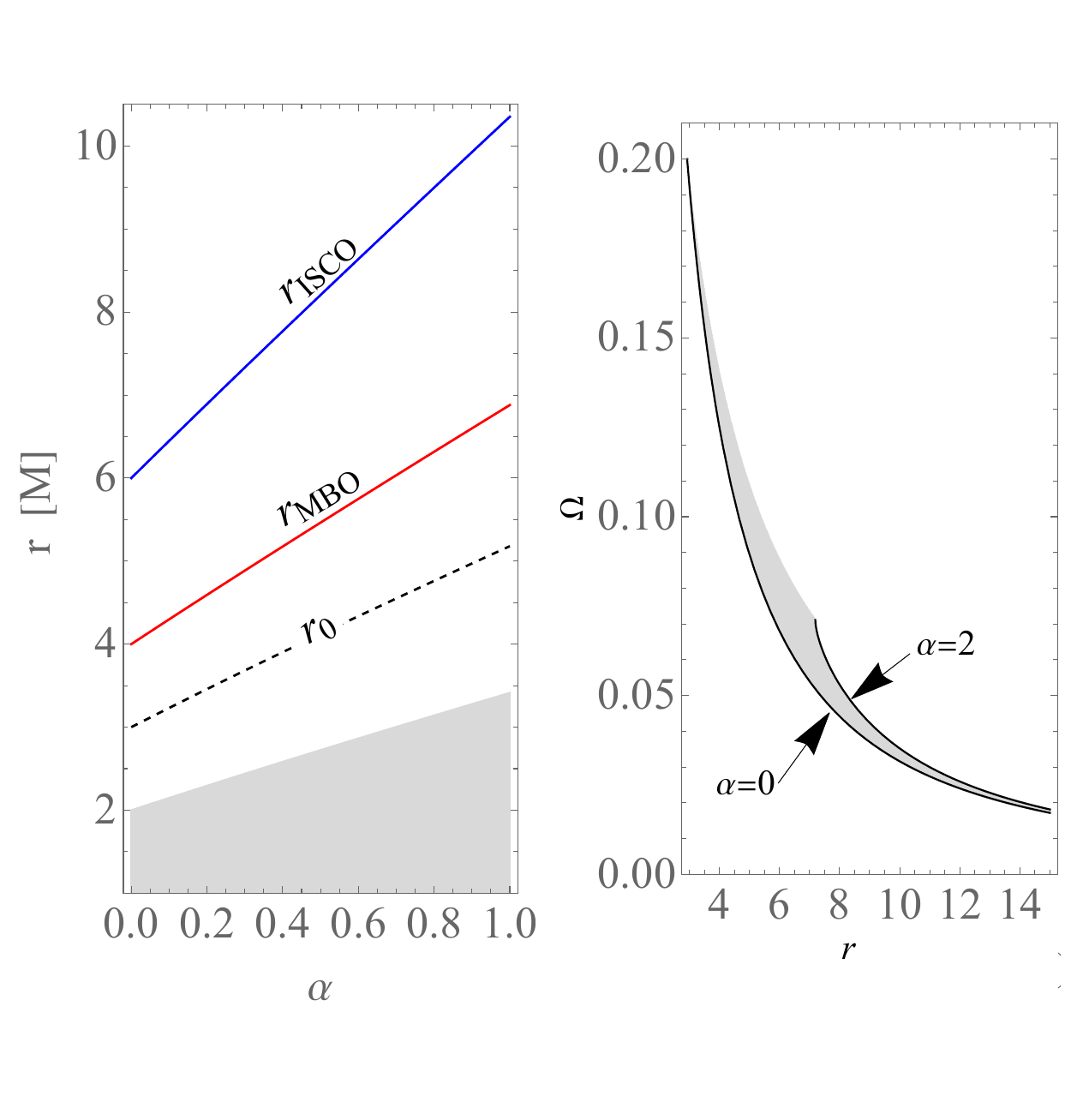}
\caption{Left panel: The ISCO and MBO positions for massive particle in the {Schwarzschild-MOG} spacetime as function of MOG parameter $\alpha$, dashed line represents the critical distance $r_0$ that angular velocity occurs, while a filled gray region represents horizon of the black hole $r_{H}^+/M$. Right panel: Radial dependence of angular velocity for different values of MOG parameter.\label{radii}}
\end{figure}
\begin{figure*}
\includegraphics[scale=0.45]{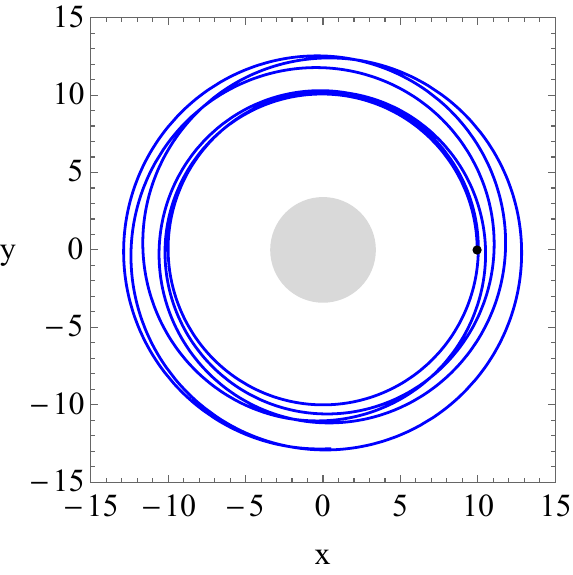}
\includegraphics[scale=0.45]{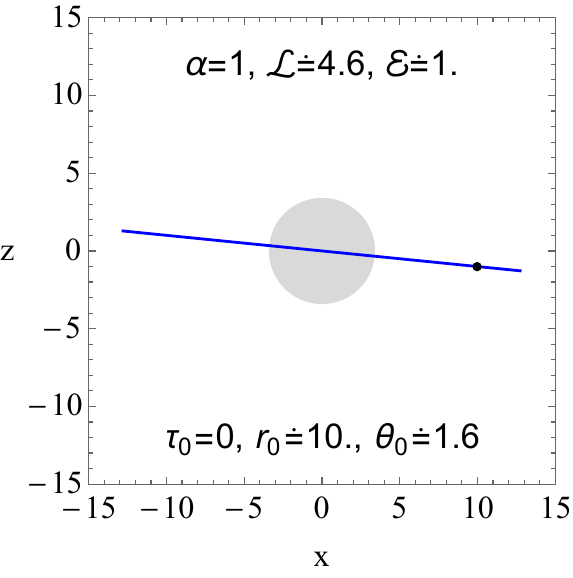}
\includegraphics[scale=0.43]{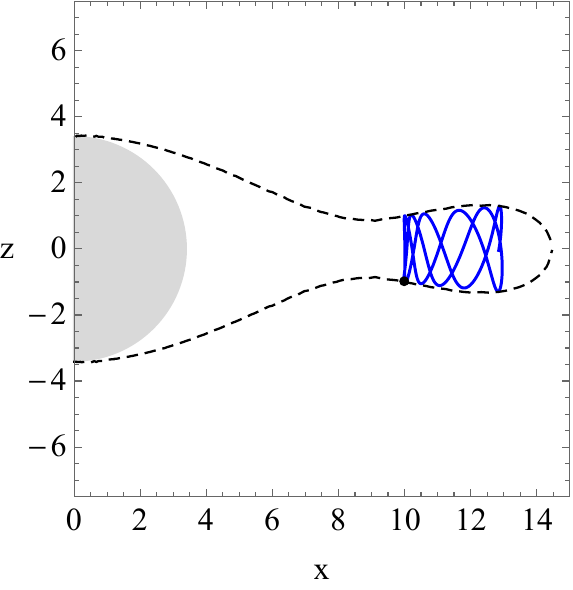}
\includegraphics[scale=0.42]{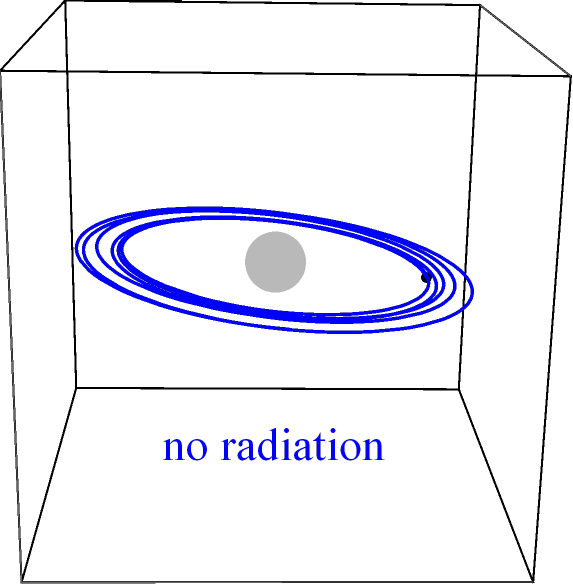}
\\
\includegraphics[scale=0.45]{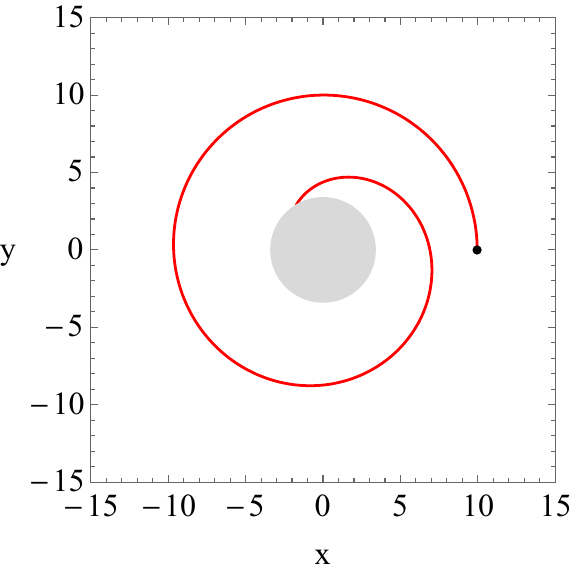}
\includegraphics[scale=0.45]{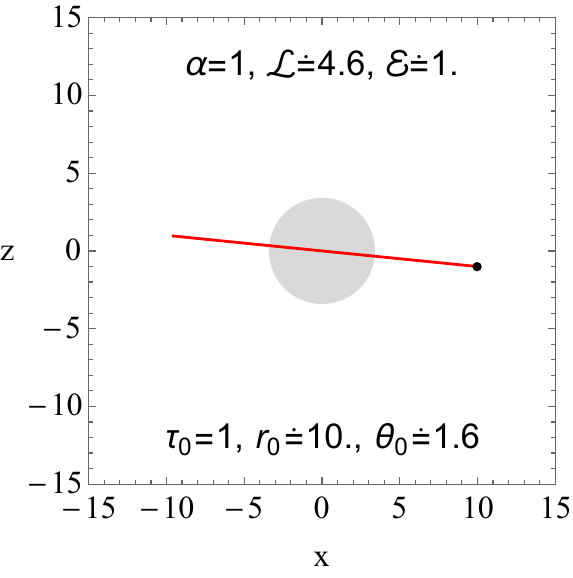}
\includegraphics[scale=0.43]{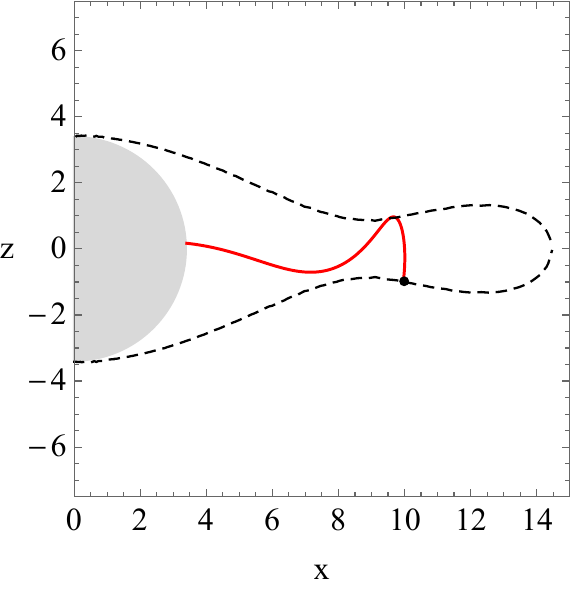}
\includegraphics[scale=0.42]{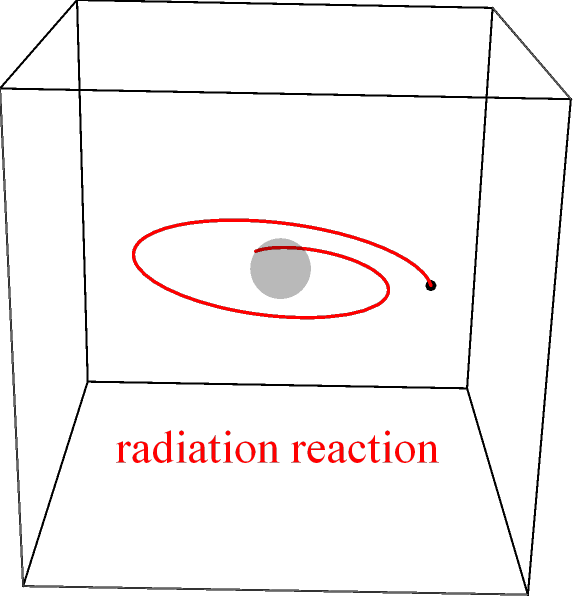}
\caption{The trajectories of massive particle orbiting the Schwarschild-MOG black hole with (bottom panel) and without (top panel) including radiation reaction force, respectively. Initial conditions are chosen to be the same for each cases and shown in the second row. The initial position of particle is depicted by black dot in each plots. The first and second columns represent the trajectory of massive particle in (x-y) and (x-z) planes, respectively. The third column represents finite motion of test particle in the region given by dashed line (contour plot of the effective potential in \eqref{Veff}), while in the last column particle's trajectory is shown in $3D$.  \label{trajectory}}
\end{figure*}

The concept of the effective potential stands as an invaluable tool for gaining insights into the trajectories of test particles in the vicinity of black holes. {It is well-known that a stationary point of the effective potential corresponds to a position of so called innermost stable circular orbit (ISCO).} Here, we focus on determining the ISCO radius for massive particles around the Schwarzschild-MOG black hole and briefly examining its dependence on MOG parameter $\alpha$. By applying the conditions $V_+(r)=\cal E$, $V_+'(r)=0$, and $V_+''(r)\leq 0$, one can readily pinpoint the ISCO radius. The dependence of the ISCO position on the MOG parameter is determined through numerical analysis, as shown in Fig.\ref{radii}. It shows that the ISCO position for massive particle forces to increases due to MOG. It is clear that all characteristic radii in this theory such as horizon, photonsphere as well as the ISCO position increases due to effect of the external vector field \cite{Moffat2015EPJCa}. It is also interesting to study marginally bound orbit (MBO) of massive particle in the Schwarzschild-MOG spacetime. The energy of particle in this orbit suppose to be equal to the rest energy, that means the critical value of the specific energy satisfies the following condition ${\cal E}=1$. Notice that in the Schwarzschild spacetime MBO radius is located at $r_{\rm MBO}=4M$. However, in MOG it depends on the $\alpha$ parameter. After careful numerical analyses we have shown dependence of the MBO position from the $\alpha$ parameter in Fig.\ref{radii} along the ISCO position. 

It is also important to discuss about angular velocity of massive particle in circular motion around the black hole. Using equation \eqref{eom} and considering motion in the equatorial plane, the angular velocity of massive particle in the Schwarzschild-MOG spacetime can be derived as \cite{Turimov2022MNRAS}
\begin{align}\label{Omega}
&\Omega=\Omega_K\sqrt{1+\alpha-\alpha\left(1+\frac{3\alpha}{2}\right)\frac{M}{r}-\alpha\sqrt{1-\frac{3M}{r}K}}\ ,  \\\nonumber
&K=1+\alpha-\frac{(8+9\alpha)M}{12r}\ ,
\end{align}
where $\Omega_K=\sqrt{M/r^3}$ is the Keplerian angular velocity. It is essential to highlight that equation \eqref{Omega} yields real results under conditions: when $K\leq r/3M$ or when $r\geq r_0$. Here, $r_0$ is determined by the expression $r_0=M(3+3\alpha+\sqrt{9+10\alpha})/2$. This signifies that the angular velocity of a massive particle occurs at a distance greater than the critical radius. In Figure \ref{radii}, we show the radial dependence of the angular velocity for various values of the MOG parameter. Additionally, the left panel of Figure \ref{radii} features the critical radius $r_0$ represented by a dashed line.

\section{Radiation reaction}\label{Sec:2}
 
Now we examine the dynamics of massive particles within the framework of the STVG theory, which incorporates a radiation reaction term. While the Lorentz-Abraham-Dirac (LAD) equation is widely recognized as the equation governing radiation reaction for charged particles {(See e.g. \cite{Poisson2004LRR,Tursunov2018ApJ})}, our research focuses on the gravitational counterpart of this phenomenon. In this context, the equation of motion \eqref{eom} can be modified as
\begin{align}\label{EQ}\nonumber
\frac{Du^\mu}{ds}&=\frac{\tilde q}{m}B^\mu_{~\nu}u^\nu+\frac{1}{2}\tau_0\left(R^\mu_{~\nu}+u^\mu u_\lambda R^\lambda_{~\nu}\right)u^\nu
\\&+\tau_0\left(\frac{D^2u^\mu}{ds^2}+u^\mu u_\lambda\frac{D^2u^\lambda}{ds^2}\right)\ , 
\end{align}
where $R_{\mu\nu}$ is the Ricci tensor, $\tau_0$ is the damping time of gravitational radiation defined as $\tau_0=2{\tilde q}^2/(3m)<<1s$ and this parameter can play a role of expansion parameter. To have an idea one can estimate of damping time for electron and it yields
\begin{align}
\tau_0=\frac{2\alpha G_N m_e}{3 c^3} \left(\frac{m}{m_e}\right) \sim 10^{-66} \alpha\left(\frac{m}{m_e}\right) {\rm s}\ ,\label{damping time}
\end{align}
while the stellar black hole (SBH) orbiting around supermassive black hole (SMBH) is
\begin{align}\label{tau}
\tau_0\simeq 3.3\times 10^{-4} \alpha\left(\frac{m}{10M_\odot}\right) {\rm s}\ .
\end{align}

Given that the final term of equation \eqref{EQ} is significantly smaller when compared to the other terms, therefore one can employ the Landau trick to expand the equation in terms of the damping time $\tau_0$. By differentiating equation \eqref{EQ} and neglecting the highest-order terms with respect to the damping time $\tau_0$, we obtain:
\begin{align}\label{Expand}
\frac{D^2u^\mu}{ds^2}&\simeq\frac{\tilde q}{m}\left(u^\alpha\nabla_\alpha B^\mu_{~\nu}+\frac{\tilde q}{m}B^\mu_{~\alpha}B^\alpha_{~\nu}\right)u^\nu+{\cal O}\left(\tau_0\right)\ .
\end{align}
Substituting expression \eqref{Expand} into to \eqref{EQ} and after simple algebraic manipulations, equation of motion can be rewritten as
\begin{align}\label{Rad}\nonumber
\frac{Du^\mu}{ds}&=\frac{\tilde q}{m}B^\mu_{~\nu}u^\nu+ \frac{1}{2}\tau_0 h^{\mu\lambda}R_{\lambda\nu}u^\nu\\&+\tau_0\frac{\tilde q}{m}\left(u^\alpha\nabla_\alpha B_{~\nu}^\mu+\frac{\tilde q}{m}h^{\mu\lambda}B_{\lambda\alpha}B^\alpha_{~\nu}\right)u^\nu\ ,
\end{align}
where $h^{\mu\nu}=g^{\mu\nu}+u^\mu u^\nu$ and $h_{\mu\nu}u^\nu=0$.

The main idea of the reducing equation \eqref{EQ} into form of \eqref{Rad} is that it is the third order system of differential equations for four coordinates. It turns out that one has to find twelve constants of motion which one of main issue in particle motion in curved spacetime. However, this problem can be avoided using the Landau trick and equation of motion reduces to the second order system of equations for four coordinates $x^\mu$. The analytical form equations for each coordinates in \eqref{Rad} are too long therefore we will not report them in the Letter. However, careful numerical analyses for given initial conditions $(0,r_0,\theta_0,0)$ and $\left\{-{\cal E}/f(r_0),0,0,{\cal L}/(r_0\sin\theta_0)^2\right\}$, allows to find each coordinates as function of affine parameter, i.e., $x^\mu=x^\mu(s)$. Hereafter performing coordinate transformation $x=r\cos\phi\sin\theta$, $y=r\sin\phi\sin\theta$, and $z=r\cos\theta$ the trajectories of particle can be produced in Cartesian coordinates using parametric plots in $2D$ or $3D$.      

\begin{figure}
\includegraphics[scale=0.6]{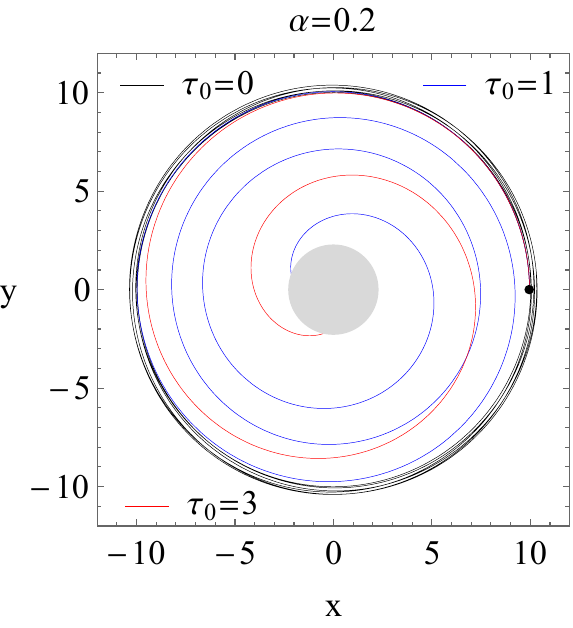}\caption{Plot is comparing the trajectories of a particle without radiation reaction force (black circular line) and with radiation reaction force for different $\tau$ parameter (blue and red). In each cases, initial positions are chosen to be the same and depicted by black dot. \label{trajectory2}}
\end{figure}

In Fig~\ref{trajectory}, we present trajectory of massive particle including and without including radiation reaction term in STVG. In order to see the effect of radiation reaction term in circular motion we set damping time to $\tau_0=1{\rm s}$. However, real value of the damping time $\tau_0$ can be calculated by following formulae \eqref{damping time} for electron and \eqref{tau} for stellar black hole. This means that damping time is much lower than $1{\rm s}$ and  particle rounds many times than shown in Fig~\ref{trajectory} around black hole and eventually fall into black hole due to radiation reaction force. From Fig.~\ref{trajectory2} one can conclude that for the larger values of MOG parameter $\alpha$ and damping time $\tau_0$, a particle falls into black hole faster.

\section{Conclusions and Discussions}\label{Sec:3}

As previously discussed, massive particles exhibit non-geodesic behavior when navigating the vicinity of black holes within the framework of STVG. This peculiar behavior arises from the similarity between the equations of motion governed by the external vector field and the Lorentz-like equation. Taking into account this fact, this research endeavors to explore the circular motion of massive test particles in the equatorial plane around the Schwarzschild-MOG black hole. We have concentrated on circular orbits and derived the effective potential that dictates the motion of these massive particles orbiting the Schwarzschild-MOG black hole. Our numerical computations have unveiled a noteworthy revelation: the ISCO position for massive particles orbiting the black hole experiences a notable shift in position due to the presence of an external vector field within the STVG framework. Furthermore, we have visually depicted this ISCO's dependence on the MOG parameter $\alpha$ through graphical representations. These findings collectively contribute to a deeper understanding of the intricate dynamics governing massive particle motion in the context of STVG theory as well as on their trajectories.

It is well-known that the Keplerian velocity is measurable quantity for distant observers in the field of astrophysics. Consequently, it becomes intriguing and significant to assess the STVG theory by examining the angular velocity of a massive particle orbiting a black hole. We have explored the angular velocity of massive particles orbiting a Schwarzschild-MOG black hole, comparing it to the Keplerian angular velocity predicted by general relativity. Additionally, we have investigated how the MOG (Modified Gravity) parameter influences angular velocity. An intriguing discovery has emerged from this research: in MOG angular velocity does not always exist itself but emerges for specific distance from the central object. This finding suggests that STVG could potentially have implications for Keplerian motion, opening up exciting avenues for further investigation in this field. Furthermore, this work offers valuable opportunities for future research endeavors, particularly in the realm of quasi-periodic oscillations (QPOs) of test particles.

In our prior publication \cite{TURIMOV2023PLB}, we demonstrated that the radiative intensity emitted by a typical stellar black hole with a mass of $10M_\odot$, orbiting around a typical supermassive black hole with a mass of $10^9 M_\odot$, can be estimated as {$I\sim 2.43\times 10^{39} {\rm erg/s}$}. To expand upon our research, we leveraged the concept that massive particles experience four-acceleration due to external vector fields, subsequently incorporating the radiation reaction term into the equation of motion. Based on this insight, we delved into the gravitational analogue of radiation reaction experienced by a massive particle orbiting a Schwarzschild-MOG black hole. By employing the Landau trick, we simplified the equation of motion by discarding high-order terms concerning damping time. Furthermore, we investigated the gravitational counterpart of radiation reaction occurring when a stellar black hole orbits a supermassive black hole. This exceptionally high intensity value suggests that such systems may serve as potent sources of gravitational waves. Indeed, the detection of gravitational waves stemming from binary black hole mergers by the LIGO and Virgo collaborations has ushered in a fresh perspective on the study of black holes and their surrounding environments within the framework of STVG. The potential observation of gravitational waves originating from other black hole systems, particularly those involving orbiting massive particles, holds the promise of yielding deeper insights into the fundamental physics governing black holes and their intricate interactions with their surroundings.

\section*{Acknowledgments}
This research was supported by the Grants F-FA-2021-510 from the Uzbekistan Ministry for Innovative Development. B.T. is grateful to prof. Moffat and Dr. Kolo{\v s} for useful comments and discussions. 

\bibliographystyle{model}  
\bibliography{Refs}

\begin{thebibliography}{44}
\expandafter\ifx\csname natexlab\endcsname\relax\def\natexlab#1{#1}\fi
\providecommand{\bibinfo}[2]{#2}
\ifx\xfnm\relax \def\xfnm[#1]{\unskip,\space#1}\fi
%Type = Article
\bibitem[{{Jackson} and {Fox}(1999)}]{Jackson1999AJP}
\bibinfo{author}{J.~D. {Jackson}}, \bibinfo{author}{R.~F. {Fox}},
\newblock \bibinfo{title}{{Classical Electrodynamics, 3rd ed.}},
\newblock \bibinfo{journal}{American Journal of Physics} \bibinfo{volume}{67}
  (\bibinfo{year}{1999}) \bibinfo{pages}{841--842}.
%Type = Article
\bibitem[{{Dirac}(1938)}]{Dirac1938RSPSA}
\bibinfo{author}{P.~A.~M. {Dirac}},
\newblock \bibinfo{title}{{Classical Theory of Radiating Electrons}},
\newblock \bibinfo{journal}{Proceedings of the Royal Society of London Series
  A} \bibinfo{volume}{167} (\bibinfo{year}{1938}) \bibinfo{pages}{148--169}.
%Type = Book
\bibitem[{Landau and Lifshitz(2004)}]{Landau-Lifshitz2}
\bibinfo{author}{L.~D. Landau}, \bibinfo{author}{E.~M. Lifshitz},
  \bibinfo{title}{The Classical Theory of Fields, Course of Theoretical
  Physics, Volume 2}, \bibinfo{publisher}{Elsevier Butterworth-Heinemann},
  \bibinfo{address}{Oxford}, \bibinfo{year}{2004}.
%Type = Article
\bibitem[{{Tursunov} et~al.(2018{\natexlab{a}}){Tursunov}, {Kolo{\v{s}}},
  {Stuchl{\'\i}k}, and {Gal'tsov}}]{Tursunov2018ApJ}
\bibinfo{author}{A.~{Tursunov}}, \bibinfo{author}{M.~{Kolo{\v{s}}}},
  \bibinfo{author}{Z.~{Stuchl{\'\i}k}}, \bibinfo{author}{D.~V. {Gal'tsov}},
\newblock \bibinfo{title}{{Radiation Reaction of Charged Particles Orbiting a
  Magnetized Schwarzschild Black Hole}},
\newblock \bibinfo{journal}{ApJ} \bibinfo{volume}{861}
  (\bibinfo{year}{2018}{\natexlab{a}}) \bibinfo{pages}{2}.
%Type = Article
\bibitem[{{Tursunov} et~al.(2018{\natexlab{b}}){Tursunov}, {Kolo{\v{s}}}, and
  {Stuchl{\'\i}k}}]{Tursunov2018AN}
\bibinfo{author}{A.~A. {Tursunov}}, \bibinfo{author}{M.~{Kolo{\v{s}}}},
  \bibinfo{author}{Z.~{Stuchl{\'\i}k}},
\newblock \bibinfo{title}{{Orbital widening due to radiation reaction around a
  magnetized black hole}},
\newblock \bibinfo{journal}{Astronomische Nachrichten} \bibinfo{volume}{339}
  (\bibinfo{year}{2018}{\natexlab{b}}) \bibinfo{pages}{341--346}.
%Type = Article
\bibitem[{{Shoom}(2015)}]{Shoom2015PRD}
\bibinfo{author}{A.~A. {Shoom}},
\newblock \bibinfo{title}{{Synchrotron radiation from a weakly magnetized
  Schwarzschild black hole}},
\newblock \bibinfo{journal}{Phys. Rev. D} \bibinfo{volume}{92}
  (\bibinfo{year}{2015}) \bibinfo{pages}{124066}.
%Type = Article
\bibitem[{{Damour} and {Deruelle}(1981)}]{Damour1981PLA}
\bibinfo{author}{T.~{Damour}}, \bibinfo{author}{N.~{Deruelle}},
\newblock \bibinfo{title}{{Radiation reaction and angular momentum loss in
  small angle gravitational scattering}},
\newblock \bibinfo{journal}{Physics Letters A} \bibinfo{volume}{87}
  (\bibinfo{year}{1981}) \bibinfo{pages}{81--84}.
%Type = Article
\bibitem[{{Damour}(1983)}]{Damour1983PRL}
\bibinfo{author}{T.~{Damour}},
\newblock \bibinfo{title}{{Gravitational Radiation Reaction in the Binary
  Pulsar and the Quadrupole-Formula Controversy}},
\newblock \bibinfo{journal}{Phys. Rev. Lett.} \bibinfo{volume}{51}
  (\bibinfo{year}{1983}) \bibinfo{pages}{1019--1021}.
%Type = Article
\bibitem[{{Blanchet} and {Damour}(1986)}]{Blanchet1986RSPSA}
\bibinfo{author}{L.~{Blanchet}}, \bibinfo{author}{T.~{Damour}},
\newblock \bibinfo{title}{{Radiative gravitational fields in general
  relativity. I. General structure of the field outside the source.}},
\newblock \bibinfo{journal}{Proceedings of the Royal Society of London Series
  A} \bibinfo{volume}{320} (\bibinfo{year}{1986}) \bibinfo{pages}{379--430}.
%Type = Article
\bibitem[{{Cutler} et~al.(1994){Cutler}, {Kennefick}, and
  {Poisson}}]{Cutler1994PRD}
\bibinfo{author}{C.~{Cutler}}, \bibinfo{author}{D.~{Kennefick}},
  \bibinfo{author}{E.~{Poisson}},
\newblock \bibinfo{title}{{Gravitational radiation reaction for bound motion
  around a Schwarzschild black hole}},
\newblock \bibinfo{journal}{Phys. Rev. D} \bibinfo{volume}{50}
  (\bibinfo{year}{1994}) \bibinfo{pages}{3816--3835}.
%Type = Article
\bibitem[{Turimov et~al.(2023)Turimov, Alibekov, Tadjimuratov, and
  Abdujabbarov}]{TURIMOV2023PLB}
\bibinfo{author}{B.~Turimov}, \bibinfo{author}{H.~Alibekov},
  \bibinfo{author}{P.~Tadjimuratov}, \bibinfo{author}{A.~Abdujabbarov},
\newblock \bibinfo{title}{Gravitational synchrotron radiation and penrose
  process in stvg theory},
\newblock \bibinfo{journal}{Phys. Lett. B} \bibinfo{volume}{843}
  (\bibinfo{year}{2023}) \bibinfo{pages}{138040}.
%Type = Article
\bibitem[{{Moffat}(2006)}]{Moffat2006JCAP}
\bibinfo{author}{J.~W. {Moffat}},
\newblock \bibinfo{title}{{Scalar tensor vector gravity theory}},
\newblock \bibinfo{journal}{JCAP} \bibinfo{volume}{2006} (\bibinfo{year}{2006})
  \bibinfo{pages}{004}.
%Type = Article
\bibitem[{{Moffat}(2015)}]{Moffat2015EPJCa}
\bibinfo{author}{J.~W. {Moffat}},
\newblock \bibinfo{title}{{Black holes in modified gravity (MOG)}},
\newblock \bibinfo{journal}{European Physical Journal C} \bibinfo{volume}{75}
  (\bibinfo{year}{2015}) \bibinfo{pages}{175}.
%Type = Article
\bibitem[{{Moffat}(2006)}]{Moffat2006CQG}
\bibinfo{author}{J.~W. {Moffat}},
\newblock \bibinfo{title}{{Time delay predictions in a modified gravity
  theory}},
\newblock \bibinfo{journal}{Classical and Quantum Gravity} \bibinfo{volume}{23}
  (\bibinfo{year}{2006}) \bibinfo{pages}{6767--6771}.
%Type = Article
\bibitem[{{Moffat} and {Toth}(2008)}]{Moffat2008ApJ}
\bibinfo{author}{J.~W. {Moffat}}, \bibinfo{author}{V.~T. {Toth}},
\newblock \bibinfo{title}{{Testing Modified Gravity with Globular Cluster
  Velocity Dispersions}},
\newblock \bibinfo{journal}{ApJ} \bibinfo{volume}{680} (\bibinfo{year}{2008})
  \bibinfo{pages}{1158--1161}.
%Type = Article
\bibitem[{{Brownstein} and {Moffat}(2007)}]{Brownstein2007MNRAS}
\bibinfo{author}{J.~R. {Brownstein}}, \bibinfo{author}{J.~W. {Moffat}},
\newblock \bibinfo{title}{{The Bullet Cluster 1E0657-558 evidence shows
  modified gravity in the absence of dark matter}},
\newblock \bibinfo{journal}{MNRAS} \bibinfo{volume}{382} (\bibinfo{year}{2007})
  \bibinfo{pages}{29--47}.
%Type = Article
\bibitem[{{Moffat}(2007)}]{Moffat2007IJMPD}
\bibinfo{author}{J.~W. {Moffat}},
\newblock \bibinfo{title}{{a Modified Gravity and its Consequences for the
  Solar System, Astrophysics and Cosmology}},
\newblock \bibinfo{journal}{International Journal of Modern Physics D}
  \bibinfo{volume}{16} (\bibinfo{year}{2007}) \bibinfo{pages}{2075--2090}.
%Type = Article
\bibitem[{{Moffat} and {Toth}(2009)}]{Moffat2009MNRASa}
\bibinfo{author}{J.~W. {Moffat}}, \bibinfo{author}{V.~T. {Toth}},
\newblock \bibinfo{title}{{The bending of light and lensing in modified
  gravity}},
\newblock \bibinfo{journal}{MNRAS} \bibinfo{volume}{397} (\bibinfo{year}{2009})
  \bibinfo{pages}{1885--1892}.
%Type = Article
\bibitem[{{Rahvar} and {Moffat}(2019)}]{Rahvar2019MNRAS}
\bibinfo{author}{S.~{Rahvar}}, \bibinfo{author}{J.~W. {Moffat}},
\newblock \bibinfo{title}{{Propagation of electromagnetic waves in MOG:
  gravitational lensing}},
\newblock \bibinfo{journal}{MNRAS} \bibinfo{volume}{482} (\bibinfo{year}{2019})
  \bibinfo{pages}{4514--4518}.
%Type = Article
\bibitem[{{Moffat} et~al.(2018){Moffat}, {Rahvar}, and
  {Toth}}]{Moffat2018Galaxies}
\bibinfo{author}{J.~{Moffat}}, \bibinfo{author}{S.~{Rahvar}},
  \bibinfo{author}{V.~{Toth}},
\newblock \bibinfo{title}{{Applying MOG to Lensing: Einstein Rings, Abell 520
  and the Bullet Cluster}},
\newblock \bibinfo{journal}{Galaxies} \bibinfo{volume}{6}
  (\bibinfo{year}{2018}) \bibinfo{pages}{43}.
%Type = Article
\bibitem[{{{\"O}vg{\"u}n} et~al.(2019){{\"O}vg{\"u}n}, {Sakall{\i}}, and
  {Saavedra}}]{Ovgun2019AP}
\bibinfo{author}{A.~{{\"O}vg{\"u}n}}, \bibinfo{author}{{\.I}.~{Sakall{\i}}},
  \bibinfo{author}{J.~{Saavedra}},
\newblock \bibinfo{title}{{Weak gravitational lensing by Kerr-MOG black hole
  and Gauss-Bonnet theorem}},
\newblock \bibinfo{journal}{Annals of Physics} \bibinfo{volume}{411}
  (\bibinfo{year}{2019}) \bibinfo{pages}{167978}.
%Type = Article
\bibitem[{{Tuleganova} et~al.(2020){Tuleganova}, {Izmailov}, {Karimov},
  {Potapov}, and {Nandi}}]{Tuleganova2020GRG}
\bibinfo{author}{G.~Y. {Tuleganova}}, \bibinfo{author}{R.~N. {Izmailov}},
  \bibinfo{author}{R.~K. {Karimov}}, \bibinfo{author}{A.~A. {Potapov}},
  \bibinfo{author}{K.~K. {Nandi}},
\newblock \bibinfo{title}{{Times of arrival (TOA) of signals in the Kerr-MOG
  black hole}},
\newblock \bibinfo{journal}{General Relativity and Gravitation}
  \bibinfo{volume}{52} (\bibinfo{year}{2020}) \bibinfo{pages}{31}.
%Type = Article
\bibitem[{{Izmailov} et~al.(2019){Izmailov}, {Karimov}, {Zhdanov}, and
  {Nandi}}]{Izmailov2019MNRAS}
\bibinfo{author}{R.~N. {Izmailov}}, \bibinfo{author}{R.~K. {Karimov}},
  \bibinfo{author}{E.~R. {Zhdanov}}, \bibinfo{author}{K.~K. {Nandi}},
\newblock \bibinfo{title}{{Modified gravity black hole lensing observables in
  weak and strong field of gravity}},
\newblock \bibinfo{journal}{MNRAS} \bibinfo{volume}{483} (\bibinfo{year}{2019})
  \bibinfo{pages}{3754--3761}.
%Type = Article
\bibitem[{{Moffat}(2015)}]{Moffat2015EPJC}
\bibinfo{author}{J.~W. {Moffat}},
\newblock \bibinfo{title}{{Modified gravity black holes and their observable
  shadows}},
\newblock \bibinfo{journal}{European Physical Journal C} \bibinfo{volume}{75}
  (\bibinfo{year}{2015}) \bibinfo{pages}{130}.
%Type = Article
\bibitem[{{Moffat} and {Toth}(2020)}]{Moffat2020PRD}
\bibinfo{author}{J.~W. {Moffat}}, \bibinfo{author}{V.~T. {Toth}},
\newblock \bibinfo{title}{{Masses and shadows of the black holes Sagittarius A*
  and M87* in modified gravity}},
\newblock \bibinfo{journal}{Phys. Rev. D} \bibinfo{volume}{101}
  (\bibinfo{year}{2020}) \bibinfo{pages}{024014}.
%Type = Article
\bibitem[{{Kolo{\v{s}}} et~al.(2020){Kolo{\v{s}}}, {Shahzadi}, and
  {Stuchl{\'\i}k}}]{Kolos2020EPJC}
\bibinfo{author}{M.~{Kolo{\v{s}}}}, \bibinfo{author}{M.~{Shahzadi}},
  \bibinfo{author}{Z.~{Stuchl{\'\i}k}},
\newblock \bibinfo{title}{{Quasi-periodic oscillations around Kerr-MOG black
  holes}},
\newblock \bibinfo{journal}{European Physical Journal C} \bibinfo{volume}{80}
  (\bibinfo{year}{2020}) \bibinfo{pages}{133}.
%Type = Article
\bibitem[{{Pradhan}(2019)}]{Pradhan2019EPJC}
\bibinfo{author}{P.~{Pradhan}},
\newblock \bibinfo{title}{{Study of energy extraction and epicyclic frequencies
  in Kerr-MOG (modified gravity) black hole}},
\newblock \bibinfo{journal}{European Physical Journal C} \bibinfo{volume}{79}
  (\bibinfo{year}{2019}) \bibinfo{pages}{401}.
%Type = Article
\bibitem[{{Rayimbaev} et~al.(2021){Rayimbaev}, {Tadjimuratov}, {Abdujabbarov},
  {Ahmedov}, and {Khudoyberdieva}}]{Rayimbaev2021Gal}
\bibinfo{author}{J.~{Rayimbaev}}, \bibinfo{author}{P.~{Tadjimuratov}},
  \bibinfo{author}{A.~{Abdujabbarov}}, \bibinfo{author}{B.~{Ahmedov}},
  \bibinfo{author}{M.~{Khudoyberdieva}},
\newblock \bibinfo{title}{{Dynamics of Test Particles and Twin Peaks QPOs
  around Regular Black Holes in Modified Gravity}},
\newblock \bibinfo{journal}{Galaxies} \bibinfo{volume}{9}
  (\bibinfo{year}{2021}) \bibinfo{pages}{75}.
%Type = Inproceedings
\bibitem[{{Manfredi} et~al.(2017){Manfredi}, {Mureika}, and
  {Moffat}}]{Manfredi2017JPCS}
\bibinfo{author}{L.~{Manfredi}}, \bibinfo{author}{J.~{Mureika}},
  \bibinfo{author}{J.~{Moffat}},
\newblock \bibinfo{title}{{Quasinormal Modes of Static Modified Gravity (MOG)
  Black Holes}},
\newblock in: \bibinfo{booktitle}{Journal of Physics Conference Series}, volume
  \bibinfo{volume}{942} of \textit{\bibinfo{series}{Journal of Physics
  Conference Series}}, p. \bibinfo{pages}{012014}.
%Type = Article
\bibitem[{{Manfredi} et~al.(2018){Manfredi}, {Mureika}, and
  {Moffat}}]{Manfredi2018PLB}
\bibinfo{author}{L.~{Manfredi}}, \bibinfo{author}{J.~{Mureika}},
  \bibinfo{author}{J.~{Moffat}},
\newblock \bibinfo{title}{{Quasinormal modes of modified gravity (MOG) black
  holes}},
\newblock \bibinfo{journal}{Physics Letters B} \bibinfo{volume}{779}
  (\bibinfo{year}{2018}) \bibinfo{pages}{492--497}.
%Type = Article
\bibitem[{{Manfredi} et~al.(2019){Manfredi}, {Mureika}, and
  {Moffat}}]{Manfredi2019JURP}
\bibinfo{author}{L.~{Manfredi}}, \bibinfo{author}{J.~{Mureika}},
  \bibinfo{author}{J.~{Moffat}},
\newblock \bibinfo{title}{{Quasinormal Modes of Modified Gravity (MOG) Black
  Holes}},
\newblock \bibinfo{journal}{Journal of Undergraduate Reports in Physics}
  \bibinfo{volume}{29} (\bibinfo{year}{2019}) \bibinfo{pages}{100006}.
%Type = Article
\bibitem[{{Moffat} and {Toth}(2019)}]{Moffat2019MNRAS}
\bibinfo{author}{J.~W. {Moffat}}, \bibinfo{author}{V.~T. {Toth}},
\newblock \bibinfo{title}{{NGC 1052-DF2 and modified gravity (MOG) without dark
  matter}},
\newblock \bibinfo{journal}{MNRAS} \bibinfo{volume}{482} (\bibinfo{year}{2019})
  \bibinfo{pages}{L1--L3}.
%Type = Article
\bibitem[{{Green} and {Moffat}(2019)}]{Green2019PDU}
\bibinfo{author}{M.~A. {Green}}, \bibinfo{author}{J.~W. {Moffat}},
\newblock \bibinfo{title}{{Modified Gravity (MOG) fits to observed radial
  acceleration of SPARC galaxies}},
\newblock \bibinfo{journal}{Physics of the Dark Universe} \bibinfo{volume}{25}
  (\bibinfo{year}{2019}) \bibinfo{pages}{100323}.
%Type = Article
\bibitem[{{Sharif} and {Shahzadi}(2017)}]{Sharif2017EPJC}
\bibinfo{author}{M.~{Sharif}}, \bibinfo{author}{M.~{Shahzadi}},
\newblock \bibinfo{title}{{Particle dynamics near Kerr-MOG black hole}},
\newblock \bibinfo{journal}{European Physical Journal C} \bibinfo{volume}{77}
  (\bibinfo{year}{2017}) \bibinfo{pages}{363}.
%Type = Article
\bibitem[{{Lee} and {Han}(2017)}]{Lee2017EPJC}
\bibinfo{author}{H.-C. {Lee}}, \bibinfo{author}{Y.-J. {Han}},
\newblock \bibinfo{title}{{Innermost stable circular orbit of Kerr-MOG black
  hole}},
\newblock \bibinfo{journal}{European Physical Journal C} \bibinfo{volume}{77}
  (\bibinfo{year}{2017}) \bibinfo{pages}{655}.
%Type = Article
\bibitem[{{P{\'e}rez} et~al.(2017){P{\'e}rez}, {Armengol}, and
  {Romero}}]{Perez2017PRD}
\bibinfo{author}{D.~{P{\'e}rez}}, \bibinfo{author}{F.~G.~L. {Armengol}},
  \bibinfo{author}{G.~E. {Romero}},
\newblock \bibinfo{title}{{Accretion disks around black holes in
  scalar-tensor-vector gravity}},
\newblock \bibinfo{journal}{Phys. Rev. D} \bibinfo{volume}{95}
  (\bibinfo{year}{2017}) \bibinfo{pages}{104047}.
%Type = Article
\bibitem[{{Della Monica} et~al.(2022{\natexlab{a}}){Della Monica}, {de
  Martino}, and {de Laurentis}}]{Monica2022Universe}
\bibinfo{author}{R.~{Della Monica}}, \bibinfo{author}{I.~{de Martino}},
  \bibinfo{author}{M.~{de Laurentis}},
\newblock \bibinfo{title}{{Constraining MOdified Gravity with the S2 Star}},
\newblock \bibinfo{journal}{Universe} \bibinfo{volume}{8}
  (\bibinfo{year}{2022}{\natexlab{a}}) \bibinfo{pages}{137}.
%Type = Article
\bibitem[{{Della Monica} et~al.(2022{\natexlab{b}}){Della Monica}, {de
  Martino}, and {de Laurentis}}]{Monica2022MNRAS}
\bibinfo{author}{R.~{Della Monica}}, \bibinfo{author}{I.~{de Martino}},
  \bibinfo{author}{M.~{de Laurentis}},
\newblock \bibinfo{title}{{Orbital precession of the S2 star in
  Scalar-Tensor-Vector Gravity}},
\newblock \bibinfo{journal}{MNRAS} \bibinfo{volume}{510}
  (\bibinfo{year}{2022}{\natexlab{b}}) \bibinfo{pages}{4757--4766}.
%Type = Article
\bibitem[{{Turimov}(2022)}]{Turimov2022MNRAS}
\bibinfo{author}{B.~V. {Turimov}},
\newblock \bibinfo{title}{{Comment on ``Orbital precession of the S2 star in
  scalar-tensor-vector gravity''}},
\newblock \bibinfo{journal}{MNRAS} \bibinfo{volume}{516} (\bibinfo{year}{2022})
  \bibinfo{pages}{434--436}.
%Type = Article
\bibitem[{{Della Monica} et~al.(2023){Della Monica}, {de Martino}, and {de
  Laurentis}}]{Monica2023MNRAS}
\bibinfo{author}{R.~{Della Monica}}, \bibinfo{author}{I.~{de Martino}},
  \bibinfo{author}{M.~{de Laurentis}},
\newblock \bibinfo{title}{{Response to: Comment on 'Orbital precession of the
  S2 star in scalar-tensor-vector gravity'}},
\newblock \bibinfo{journal}{MNRAS} \bibinfo{volume}{521} (\bibinfo{year}{2023})
  \bibinfo{pages}{474--477}.
%Type = Article
\bibitem[{{Capozziello} et~al.(2020){Capozziello}, {Mantica}, and
  {Molinari}}]{Capozziello2020IJMPD}
\bibinfo{author}{S.~{Capozziello}}, \bibinfo{author}{C.~A. {Mantica}},
  \bibinfo{author}{L.~G. {Molinari}},
\newblock \bibinfo{title}{{General properties of f(R) gravity vacuum
  solutions}},
\newblock \bibinfo{journal}{International Journal of Modern Physics D}
  \bibinfo{volume}{29} (\bibinfo{year}{2020}) \bibinfo{pages}{2050089}.
%Type = Article
\bibitem[{{Bajardi} et~al.(2022){Bajardi}, {D'Agostino}, {Benetti}, {De Falco},
  and {Capozziello}}]{Bajardi2022EPJP}
\bibinfo{author}{F.~{Bajardi}}, \bibinfo{author}{R.~{D'Agostino}},
  \bibinfo{author}{M.~{Benetti}}, \bibinfo{author}{V.~{De Falco}},
  \bibinfo{author}{S.~{Capozziello}},
\newblock \bibinfo{title}{{Early and late time cosmology: the f(R) gravity
  perspective}},
\newblock \bibinfo{journal}{European Physical Journal Plus}
  \bibinfo{volume}{137} (\bibinfo{year}{2022}) \bibinfo{pages}{1239}.
%Type = Article
\bibitem[{{Moffat} and {Toth}(2009)}]{Moffat2009CQG}
\bibinfo{author}{J.~W. {Moffat}}, \bibinfo{author}{V.~T. {Toth}},
\newblock \bibinfo{title}{{Fundamental parameter-free solutions in modified
  gravity}},
\newblock \bibinfo{journal}{Classical and Quantum Gravity} \bibinfo{volume}{26}
  (\bibinfo{year}{2009}) \bibinfo{pages}{085002}.
%Type = Article
\bibitem[{{Poisson}(2004)}]{Poisson2004LRR}
\bibinfo{author}{E.~{Poisson}},
\newblock \bibinfo{title}{{The Motion of Point Particles in Curved Spacetime}},
\newblock \bibinfo{journal}{Living Reviews in Relativity} \bibinfo{volume}{7}
  (\bibinfo{year}{2004}) \bibinfo{pages}{6}.

\end{thebibliography}
\end{document}